\newif\ifsubmissionversion % Short anonymized version without comments and authors
\submissionversionfalse

\newif\ifsubfolders % Short anonymized version without comments and authors
\subfoldersfalse

\documentclass[nonacm,sigconf,natbib=false]{acmart} % nonacm, suppresses the copyright box
\AtBeginDocument{%
  }

\usepackage{hyperref}
\usepackage{algorithm}
\usepackage{algpseudocode}
\usepackage{amsmath}
\usepackage{enumitem}
\usepackage{amsthm}
\usepackage{tcolorbox}
\newtheorem{definition}{Definition}
\usepackage{tikz}
\usepackage{caption}
\usepackage{cryptocode}
\usepackage{pifont}
\usepackage{adjustbox}
\usepackage{xcolor}
\usepackage[dvipsnames]{xcolor}
\usepackage{cleveref}
\crefname{section}{§}{§§}
\Crefname{section}{§}{§§}

\newcounter{myctr}
\def\fH{\ensuremath{\mathcal{H}}} % Caligraphic for hash

\def\c#1{\ensuremath{\mathtt{#1}}} % Math TT for Functions

\def\f#1{\ensuremath{\mathrm{#1}}} % Math Roman for Functions
\def\fVerify{\f{Verify}} % roman for functions

\def\fHMAC{\f{HMAC}}
\def\fEnc{\f{Enc}}

\def\fUnseal{\f{Unseal}}
\def\fRandom{\f{Random}}
\def\fHMAC{\f{HMAC}}
\def\v#1{\ensuremath{\mathit{#1}}} % Math Italic for variables
\def\vPK{\v{PK}} % Italic for variables
\def\vSK{\v{SK}} % Italic for variables
\def\vCert{\v{Cert}} % Italic for variables
\def\vMRENCLAVE{\v{MRENCLAVE}} % Italic for variables
\def\vnonce{\v{nonce}} % Italic for variables
\def\vstate{\v{state}} % Italic for variables
\def\vENC{\v{ENC}}
\def\vMAC{\v{MAC}}
\def\vpuzzle{\v{puzzle}}
\def\vMRSIGNER{\v{MRSIGNER}}

\usepackage{wasysym}

\begin{document}

\title[Reinforcing Secure Live Migration through Verifiable State Management]{Reinforcing Secure Live Migration\\ through Verifiable State Management}

\author{Stefanos Vasileaidis}
\email{stefanos.vasileiadis@unitn.it}
\orcid{0009-0005-2193-4464}
\affiliation{%
\institution{Univ. of Trento}
\country{Italy}
}
\author{Thanassis Giannetsos}
\email{agiannetsos@ubitech.eu}
\orcid{to be updated}
\affiliation{%
\institution{Ubitech}
\country{Greece}
}
\author{Matthias Schunter}
\email{schunter@acm.org}
\orcid{0000-0001-9760-2726}
\affiliation{%
\institution{Intel Labs}
\country{Germany}
}
\author{Bruno Crispo}
\email{crispo@disi.unitn.it}
\orcid{to be updated}
\affiliation{%
\institution{Univ. of Trento}
\country{Italy}
}

\begin{abstract}
Live migration of applications is a fundamental capability for enabling resilient computing in modern distributed systems. However, extending this functionality to trusted applications (TA) — executing within Trusted Execution Environments (TEEs) — introduces unique challenges such as secure state preservation, integrity verification, replay and rollback prevention, and mitigation of unauthorized cloning of TAs. We present TALOS, a lightweight framework for verifiable state management and trustworthy application migration. While our implementation is prototyped and evaluated using Intel SGX with the Gramine LibOS and RISC-V Keystone (evidencing the framework’s portability across diverse TEEs), its design is agnostic to the underlying TEE architecture. Such agility is a necessity in today's network service mesh (collaborative computing across the continuum) where application workloads must be managed across domain boundaries in a harmonized fashion. TALOS is built around the principle of minimizing trust assumptions: TAs are treated as untrusted until explicitly verified, and the migration process does not rely on a trusted third party. To ensure both the integrity and secure launch of the migrated application, TALOS integrates memory introspection and control-flow graph extraction, enabling robust verification of state continuity and execution flow. Thereby achieving strong security guarantees while maintaining efficiency, making it suitable for decentralized settings.
\end{abstract}


\maketitle

\section{Introduction}
\label{sec:intro}

Over the past decade, there has been a continuous shift towards cloud-native and edge computing paradigms, enabling customers to exploit both advanced computational capabilities and cost-effective deployment strategies. These paradigms have evolved alongside virtualization technologies to realize next-generation ``\emph{Systems-of-Systems}'' (SoS), transforming isolated, standalone service siloes into secure, distributed solutions spanning the continuum from the core cloud to the distributed edge and far-edge cyber-physical systems~\cite{Schneider_Dhar_Puddu_Kostiainen_Čapkun_2021}. Fundamental to this digital network infrastructure transformation (and a long-run objective) is the materialization of the \emph{3C Network} vision bringing forth a new cognitive, collaborative computing model~\cite{white-paper-2025} aiming at the optimal synergy between IoT and agentic AI, disrupting the boundaries of automation and autonomous cooperation: from chips and other components for high-speed processors embedded in devices, to edge computing working cohesively with centralised cloud services and AI-powered applications managing tasks delegated from human operators~\cite{hexa-x}. This allows computing to be integrated everywhere in the network.

Translating these ideas to the digital realm, we now face the challenge of delegating tasks to AI agents. Early AI systems were built on rigid, command-and-control architectures. Today, however, there’s a shift toward systems where AI agents act autonomously — making decisions, communicating with one another, and learning from their interactions. This evolution is akin to moving from an authoritarian regime to a more collaborative, distributed model of governance. This is where Dynamic Function Placement (DFP) has been described for deploying/migrating functions to orchestrate differentiated services optimally across multiple sites and infrastructures based on diverse intents and policy constraints of changing environments~\cite{white-paper-2025}. The concept of DFP unlocks \emph{service mesh} paradigms, with uniform resourcing space, where AI agents (as functions) are managed over domain boundaries in a harmonized manner. However, realizing such dynamic management and orchestration requires not only secure relocation, replication and offloading capabilities but also \emph{verifiabe} application state migration—ensuring strong security guarantees throughout the process.

%Specifically, the integrity and trustworthiness of both code and data must be provable at every stage of migration. 
Existing migration schemes typically prioritize low latency and efficiency~\cite{article-state}, often targeting entire VMs and/or containers rather than the finer granularity of application state~\cite{2017SGX}. In this context, it is found that many resources are passed across the network during migration, such as CPU state, network state, and memory state. The transfer of the whole memory state takes too long to be a realistic solution in autonomous networks where decision making needs to be distributed across agents that need to share the same perception (state)~\cite{article-2025}. This issue becomes especially problematic as
containerized programs change large amounts of memory more quickly than a container can transmit across the network, thus, limiting the usefulness of of existing live migration techniques which are based on ``checkpoint and restore'' mechanisms~\cite{live-migration}; i.e., all pre-copy, post-copy and hybrid container migration techniques while optimizing the migration cycle, they cannot cope with more agile memory management for migrating dynamically calculated volatile memory pages/states.  

A fundamental challenge compounding this issue is ensuring \emph{execution integrity prior to the transfer of state}. Hardware-assisted security mechanisms have emerged as effective solutions to this problem~\cite{asokan}, offering stronger guarantees by minimizing the Trusted Computing Base (TCB) and shifting trust towards lower layers such as firmware or immutable hardware. This paradigm has led to the development of \emph{Trusted Execution Environments} (TEEs), which provide isolated execution contexts for sensitive code and data. While TEE implementations vary, they generally enforce memory isolation from privileged software such as operating systems and hypervisors. Remote attestation protocols allow external parties to verify the authenticity and integrity of the software running inside a TEE. However, existing hardware-enforced protections are primarily optimized for container or VM-level migration and do not address the unique challenges of migrating application-level TEE state, which is often tightly bound to hardware-specific sealing mechanisms and runtime context.

This raises the question of \emph{how to monitor -- during runtime -- the behaviour of an enclave \footnote{In this work, the terms \emph{enclave} and \emph{trusted application} are used interchangeably.} so as to ensure application state integrity during and after the migration process?} Maintaining that the migrated application continues to operate securely in its new environment is critical, especially in scenarios where sensitive data and computations need to be transferred to elements that are characterized as highly trustworthy. Usually, the hardware-based secure element (e.g., SGX) has many states for each running enclave. All these states are necessary for resuming the enclave’s execution on the target machine, including the enclave memory, data structures, counter values, etc. (\cref{subsec:state}). However, neither the hypervisor nor the guest OS can access an enclave’s memory directly. If the guest operating system extracts the memory state of a trusted application, the data cannot be directly reused; instead, the state must be sealed to ensure that it can only be restored and executed within a verified and authorized TEE context. Thus, new introspection capabilities are required that balance state monitoring with the security level offered and without breaching the isolation level of an enclave.

All in all, traditional live (container or VM-oriented) migration techniques cannot address these challenges, as they lack the ability to inspect or validate the internal state of an enclave-protected application without compromising its security. This is especially problematic for long-running, stateful applications where continuity of execution semantics is critical. TALOS addresses this gap by introducing novel introspection and verification mechanisms that preserve state integrity while maintaining TEE isolation properties.

\textbf{\emph{Contributions:}} In this paper, we present \textbf{\emph{TALOS}}, a lightweight and architecture-agnostic framework for secure and verifiable migration of enclave applications with persistent state. TALOS is the first to combine verifiable state extraction, attested transfer, and Proof of Execution (PoX) under a zero-trust model—ensuring that the application’s integrity and state continuity are preserved throughout the migration lifecycle, including during relaunch on the target node. Even under an extended adversarial model (c\ref{subsec:threat_model}), TALOS guarantees that both code and data are correctly reconstructed in the destination enclave and can resume execution without violating trust assumptions. To achieve this, TALOS defines a minimal yet sufficient Trusted Computing Base (TCB), comprising only those software components necessary for controlled state extraction, measurement, and control-flow verification. These operate in concert with the TEE's hardware root of trust and are subject to remote attestation across all phases of migration. Our prototype extends the Gramine Library Operating System (LibOS)~\cite{10.1145/3658644.3690323} atop Intel SGX, leveraging its lightweight and scalable architecture for isolating unmodified applications in enclaves. Additionally, we adapt the RISC-V Keystone TEE to support TALOS’s core mechanisms—most notably, the PoX pipeline—demonstrating its agile nature.

\textbf{Design:} TALOS architecture remains TEE-agnostic (Fig.~\ref{fig:hlo}). It leverages enclave-based isolation to introspect application state and execution behavior after migration. Rather than assuming the trustworthiness of the migration process, TALOS performs explicit verification of the code loaded into enclave memory and its runtime behavior. This includes parsing the binary’s ELF structure to validate critical metadata and correlating this with dynamic control-flow traces to ensure both continuity and integrity of execution.

At the core of TALOS lies its \textbf{Proof of Execution (PoX)} mechanism, an introspection capability integrated into the Gramine-based measurement root. PoX monitors control-flow transitions in real time to detect anomalies such as code injection, unauthorized forks, or unexpected system calls. Unlike traditional attestation, which captures only static measurements, TALOS continuously verifies that execution adheres to the expected control-flow graph and originates from an unmodified, approved binary (throughout the migration cycle). This provides strong evidence that the migrated application has not been tampered with and is operating as intended—even in untrusted environments.

TALOS further strengthens this guarantee by enforcing an one-to-one mapping between application instances and their system-level identifiers, preventing replication and fork-based cloning attacks. To protect against resource exhaustion threats such as fork bombs, TALOS enforces bounded instantiation by maintaining a structured identity mapping. This limits the number of active enclave instances and ensures that redundant or malicious process spawns can be detected and blocked, preserving system stability.

Crucially, TALOS requires only minimal integration with the TEE’s attestation pipeline and does not depend on vendor-specific infrastructure or a centralized root of trust. Its modular design allows it to operate as a migration enabler and silent observer across heterogeneous TEE platforms, as long as basic security primitives—such as enclave isolation, remote attestation, and sealed state protection—are available. This flexibility is demonstrated by our custom extension of the Keystone TEE, where we adapted TALOS to support real-time PoX, showcasing its portability across architectural boundaries.

\section{Preliminaries}
\label{sec:prel}

\subsection{Trusted Computing Base (TCB)}
\label{subsec:gramine}

For a given service or function, the TCB encompasses the minimal set of hardware, firmware, and software components that must function correctly so as to uphold its security guarantees~\cite{10.1145/3555776.3577809}. TALOS relies on an intentionally small and well-defined TCB for safeguarding the migration cycle. Rather than introducing heavy, platform-dependent modifications, TALOS encompasses lightweight software extensions (based on non-intrusive state transformation utilities, removing the necessity to add a runtime into the target address space (\cref{subsec:affinity} and \cref{subsec:import}) to enforce and verify correct migration semantics across heterogeneous elements. To anchor trust and ensure the TCB remains verifiable, TALOS manifests a Root of Trust (RoT) - featuring measurement and reporting capabilities - for reliably creating (evidence-based) claims of the correct and uninterrupted instantiation and continuity of both persistent and volatile application states (\cref{subsec:state}). This mechanism ensures that the integrity and continuity of the migration can be attested, even under adversarial and dynamic threat models.

\subsection{TEEs \& Attestation Services}

\textbf{Trusted Execution Environments (TEEs)} provide hardware-backed isolation for security-critical application components. They ensure that sensitive code and data execute in a protected context that is shielded from external software, including operating systems, hypervisors, and other user-level applications. TEEs typically rely on dedicated hardware features—such as memory encryption, access control enforced by the Memory Management Unit (MMU), and tamper-resistant storage—to preserve confidentiality and integrity even in adversarial environments. TEE memory regions are cryptographically protected and inaccessible to software running outside the trusted environment. Modern TEE designs also incorporate protection against rollback and replay attacks, often using integrity trees or monotonic counters to prevent state manipulation. These guarantees form the foundation for secure execution, even on potentially compromised platforms.

\textbf{Remote attestation} is a key mechanism for establishing trust in TEEs. It enables a relying party to verify the identity and integrity of a remote enclave application before provisioning secrets or initiating sensitive interactions. Attestation typically involves generating a signed report that includes measurements of the enclave's initial state (e.g., a cryptographic hash of its code and configuration), often referred to as a measurement root. These reports are signed by a hardware-protected attestation key, verifiable either directly (as in open attestation models) or via privacy-preserving group signatures (as in anonymous attestation schemes). Some platforms also support developer signing keys or signer identities, allowing trust to be bound not only to the code itself but also to its provenance.

In general, TEEs derive enclave-specific cryptographic material—such as keys or sealed blobs—based on platform state and attested measurements. This ensures that secrets remain tightly bound to the expected execution environment and cannot be extracted or reused by malicious or unverified code.

\section{\emph{TALOS} Design Principles}

\subsection{Real-Time Proof of Execution}
\label{sec:PoX}
Ensuring that an enclave application adheres to trusted execution semantics requires more than static launch-time attestation; it demands run-time integrity guarantees. \textbf{Proof of Execution (PoX)}~\cite{PoX} extends traditional attestation by verifying that an application not only loads correctly but also behaves as expected during execution. TALOS builds upon this idea by introducing a lightweight, real-time verification framework tailored for dynamic, migration-aware enclave applications.

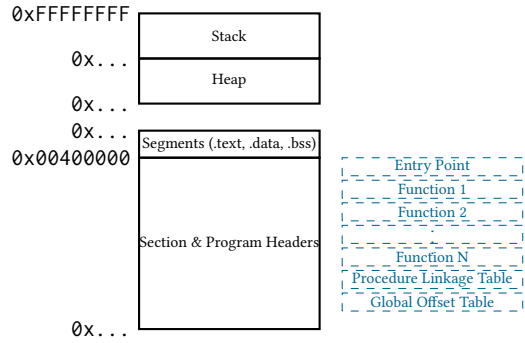
\begin{figure}[t]
    \centering
    \begin{tikzpicture}[scale=0.6]
        % Stack and Heap
        \draw[thick] (0, 6) rectangle (4, 7) node[pos=.5] {\scriptsize Stack};
        \draw[thick] (0, 5) rectangle (4, 6) node[pos=.5] {\scriptsize Heap};
        
        % Segments (smaller box)
        \draw[thick] (0, 3.8) rectangle (4, 4.4) node[pos=.5] {\scriptsize Segments (.text, .data, .bss)};
        
        % ELF & Program Headers (larger box)
        \draw[thick] (0, 0) rectangle (4, 3.8) node[pos=.5] {\scriptsize Section \& Program Headers};

        % Memory Addresses
        \node[left] at (0, 7) {\texttt{0xFFFFFFFF}};
        \node[left] at (0, 6) {\texttt{0x...}};
        \node[left] at (0, 5) {\texttt{0x...}};
        \node[left] at (0, 4.4) {\texttt{0x...}};
        \node[left] at (0, 3.8) {\texttt{0x00400000}};
        \node[left] at (0, 0) {\texttt{0x...}};

        % ELF & Function Section (More Space Between Entries)
        \color{MidnightBlue} \draw[dashed] (4.5, 3.4) rectangle (8.5, 3.8) node[pos=.5] {\scriptsize Entry Point};
        \draw[dashed] (4.5, 2.9) rectangle (8.5, 3.3) node[pos=.5] {\scriptsize Function 1};
        \draw[dashed] (4.5, 2.4) rectangle (8.5, 2.8) node[pos=.5] {\scriptsize Function 2};
        \draw[dashed] (4.5, 1.9) rectangle (8.5, 2.3) node[pos=.5] {\scriptsize \vdots};
        \draw[dashed] (4.5, 1.4) rectangle (8.5, 1.8) node[pos=.5] {\scriptsize Function N};
        \draw[dashed] (4.5, 0.9) rectangle (8.5, 1.3) node[pos=.5] {\scriptsize Procedure Linkage Table};
        \draw[dashed] (4.5, 0.4) rectangle (8.5, 0.8) node[pos=.5] {\scriptsize Global Offset Table};
    \end{tikzpicture}
    \caption{Memory Layout of an ELF Binary with Function Symbols and Metadata}
    \label{fig:elf_memory_mapping}
    \vspace{-0.3cm}
\end{figure}

Our approach begins with static verification of the enclave binary, which follows the Executable and Linkable Format (ELF)—a standard for Linux-based binaries. In ELF, the \texttt{.text} segment contains the executable code of the program, while the symbol table encodes metadata that maps symbolic names to their corresponding memory addresses, enabling linking and debugging. This involves hashing critical sections (e.g., \texttt{.text}) and validating the binary layout and symbol resolution to detect tampering or injection(Fig. ~\ref{fig:elf_memory_mapping}). Run-time behavior is monitored through \textbf{control-flow integrity (CFI)} enforced by a system call control-flow graph (SC-CFG)~\cite{10.1145/3548606.3559344}, which allows us to detect anomalous transitions caused by control-flow hijacking attacks such as return-oriented programming.

\begin{definition}
Given a program, an SC-CFG is a directed graph \( G = (V, E) \), where \( V \) is the set of system calls (i.e., nodes), and \( E \) is the set of edges representing transitions between system calls. A control flow edge from system call \( v_i \) to \( v_j \) is denoted as \( (v_i, v_j) \), where \( V \) and \( E \) are the node and edge sets of \( G \).
\end{definition}

PoX frameworks typically require a prover to execute a specific function whose output, combined with a hash of the code and a nonce, forms an attestation token. This token is only valid if the execution was atomic and unmodified. While TALOS does not adopt this strict atomicity—impractical for interactive or long-running applications—it incorporates PoX-inspired semantics to validate execution integrity post-migration. Specifically, it can verify that resumed enclave execution aligns with the original application's control-flow and memory state, without requiring hardware changes or platform-specific modifications. This accounts for non-intrusiveness and portability, relying solely on minimal, software-centric extensions to maintain trust in dynamic environments.

\subsection{State Migration}
\label{subsec:state}

Secure enclave migration hinges on two complementary components of the application’s execution context: the \textbf{persistent state}, which captures static program structure, and the \textbf{volatile state}, which encompasses dynamic runtime data. While most prior approaches verify only the persistent binary, TALOS also attests to the volatile state post-migration—ensuring the resumed execution faithfully reflects the original application's state and logic. This enables TALOS to deliver a unified verification model that reinforces the \textbf{Proof of Execution} guarantees discussed earlier.

\subsubsection{\textbf{Persistent State}}

The persistent state defines the stable structure of the application loaded into the target Trusted Execution Environment (TEE). This includes (i) the \textbf{in-memory layout} of the ELF binary—specifically sections like \texttt{.text}, \texttt{.data}, and \texttt{.bss}, along with symbol and relocation tables—and (ii) the dynamic \textbf{control-flow graph (CFG)}, derived from the sequence of system calls executed during the reinitialization phase, which constitutes the runtime behavior of the migrated trusted application prior to resuming normal execution.

TALOS validates the binary by inspecting its ELF structure, ensuring that code and data segments are correctly mapped and that symbol resolution remains consistent with the original. At runtime, it dynamically captures the control-flow transitions to confirm that the application executes the necessary logic to ingest its volatile state. This dual inspection ensures that the application is not only structurally valid but also logically sound following migration. Anomalies such as incorrect relocations, missing symbols, or deviations from expected system-call patterns are treated as signs of tampering or failure, invalidating trust in the enclave’s state.

\subsubsection{\textbf{Volatile State}}

The volatile state captures the live runtime context necessary for correct execution after migration. It includes heap and stack contents, memory allocations, open file descriptors, cryptographic keys, sealed data, and other enclave-specific resources such as monotonic counters. In TALOS, enclave applications are migration-aware: they externalize this volatile state securely before migration, allowing it to be encrypted, transferred, and restored into the new enclave instance.

This state is verified upon loading into the destination enclave before execution resumes. Even if the persistent structure is intact, any corruption or inconsistency in the volatile state may break semantic correctness or leak sensitive data. TALOS ensures that state continuity is preserved end-to-end, making enclave migration both deterministic and secure.

\section{System \& Threat Model}
\subsection{System Model}
\label{subsec:system_model}
At the heart of TALOS lies a minimal and well-defined \textbf{Trusted Computing Base (TCB)}, responsible for executing cryptographic operations and performing runtime introspection critical to the security of the migration process. This TCB operates under the principle of a \textit{Root of Trust for Measurement}, ensuring that sensitive operations—such as state extraction, transfer, and verification—are executed within an isolated and attested environment.

To fulfill these requirements, TALOS assumes the presence of a \textbf{Trusted Execution Environment (TEE)} that offers hardware-enforced isolation, secure memory regions, and privileged execution capabilities. The TEE must be sufficiently expressive—typically through OS-like abstractions—to allow TALOS components to perform introspection, control-flow tracing, and state integrity checks within the enclave boundary. In this model, the TEE itself becomes the foundation for measurement and trust, enabling verifiable execution even in the presence of an untrusted operating system or hypervisor.

The TALOS protocol is structured around these principles and relies solely on this minimal TCB to provide strong integrity guarantees throughout the migration lifecycle.

A \textbf{Migrating Node} refers to a zero-trust device equipped with a Trusted Execution Environment (TEE) and can function either as a \emph{Source Migration Node (SMN)} or a \emph{Target Migration Node (TMN)}. Each node hosts a \textbf{Migration Service Trusted Application (TA)}, which orchestrates the deployment and management of enclave applications and constitutes the core of the device’s \textbf{Trusted Computing Base (TCB)}. 

In the SMN role, the TA initiates measurement and verifiable extraction of the application's volatile state. As a TMN, it receives this state and resumes execution within a secure context, providing guarantees of \textbf{authenticity} and \textbf{integrity} at launch. The Migration Service operates entirely within the TEE, leveraging a certified key pair to preserve the \textbf{confidentiality, authenticity, and integrity} of the application state during the migration process. \textbf{Migrated applications are instantiated as child processes of the TA}—a deliberate design choice that enables secure and introspective access to the memory layout of these applications. By leveraging trusted parent-child semantics, the TA can maintain fine-grained visibility over enclave state, while respecting the isolation guarantees provided by the underlying TEE architecture.

An \textbf{Orchestrator} is a trusted entity responsible for (i) onboarding migration nodes by verifying their initial key certificates and (ii) validating the configuration of each node’s Migration Service. It serves as a trusted third party and may be implemented as a custom service or layered atop existing attestation infrastructures, such as Intel’s Attestation Service \cite{9217791, 10.1007/978-3-030-63406-3_16}. Although the presence of a trusted third party is assumed during initialization, \textbf{TALOS adopts a decentralized attestation architecture} for the migration process itself. All interaction with the Orchestrator is limited to a one-time enrollment phase during the provisioning of each Migration Service. Once this phase is complete, live migration proceeds autonomously and without further reliance on centralized trust anchors \cite{etsi2019zsm}.

\begin{figure}[!t]
    \centering
    \includegraphics[width=0.32\textwidth]{\ifsubfolders figures/\fi 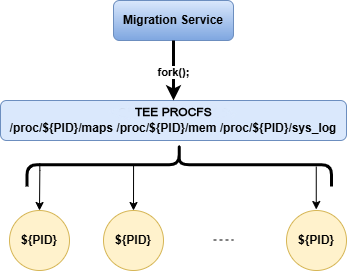}
    \caption{An Insider’s Perspective on the TMN Architecture}
    \label{fig:arch}
    \vspace{-0.2cm}
\end{figure}

\subsection{Threat Model}
\label{subsec:threat_model}

TALOS is agnostic to the underlying \textbf{Root of Trust (RoT)}, and its design is compatible with any TEE that offers hardware-enforced isolation and secure key provisioning. Our design follows the standard threat model of confidential computing~\cite{10.1145/2487726.2488368}, wherein the host software stack—including the operating system, hypervisor, and system libraries—is treated as untrusted. We assume that the adversary has full control over the host software and may possess \textbf{physical access} to the machine. The attacker can intercept, replay, or manipulate both host-level and network communications, and can arbitrarily introspect or tamper with non-enclave memory. Trust is placed solely in the \textbf{TALOS TCB} and the underlying hardware-based TEE mechanism, which together enforce enclave isolation, sealed state protection, and secure key handling. The enclave must remain resilient against attacks such as Control-Oriented Programming in Enclaves (COIN)~\cite{10.1145/3373376.3378486}, which target the integrity of the enclave’s control flow without violating memory boundaries. We assume that all cryptographic primitives are correctly implemented and cannot be broken by the adversary. Like most enclave systems, TALOS does \emph{not} defend against \textbf{side-channel attacks}, including but not limited to traditional cache timing~\cite{10.1109/SP.2015.43}, page-table based~\cite{10.5555/3241189.3241271, 10.1145/3133956.3134038, 10.1109/SP.2015.45}, or speculative execution attacks~\cite{10.1145/3399742, 8835281}. These threats are considered \emph{orthogonal} to our work and must be addressed through complementary defenses at the microarchitectural or runtime level.

% \subsubsection{\textbf{Software Adversary and Tampering Attacks}}
% 
% We assume the presence of a sophisticated software adversary capable of compromising both the source and target nodes—excluding the TEE. This adversary can mount tampering attacks on both the control and data planes of the migration process. On the control plane, the attacker may exploit vulnerabilities in migration orchestration services. For example, CVE-2024-5910 in Palo Alto Networks’ Expedition tool allows remote attackers to bypass authentication and gain administrative control, exposing all configuration data to tampering. On the data plane, the adversary may intercept or modify the application's memory state during transmission, e.g., via Man-in-the-Middle (MitM) attacks. A more severe case involves achieving a VM escape, allowing direct access to the hypervisor. For instance, the vulnerability chaithe attacker may exploit vulnerabilities in migration orchestration of CVE-2025-22224, CVE-2025-22225, and CVE-2025-22226 in VMware enables a guest VM to execute arbitrary code on the host, granting complete control over all migrating applications.
\subsubsection{\textbf{Software Adversary and Tampering Attacks}}

We assume the presence of a sophisticated software adversary capable of compromising both the source and target nodes—excluding the TEE. In this context, we distinguish two aspects of the migration process: (i) the \emph{migration-control aspect}, which encompasses orchestration services, automation tooling, and authorization components that initiate and coordinate migration; and (ii) the \emph{migration-data aspect}, which covers the serialized application state and its transmission between nodes.

On the migration-control aspect, the attacker may exploit vulnerabilities in orchestration services. For example, CVE-2024-5910 in Palo Alto Networks’ Expedition (a configuration migration tool) allows unauthenticated administrative takeover, exposing stored configurations, credentials, and templates to tampering. On the migration-data aspect, the adversary may intercept or modify the application’s serialized memory/state during transmission (e.g., via man-in-the-middle attacks), or leverage compute-boundary flaws to access state directly. Notably, recent VMware vulnerabilities (CVE-2025-22224, CVE-2025-22225, CVE-2025-22226) demonstrate guest-to-host impact—enabling code execution in the host’s VMX process and potential kernel-level compromise under certain conditions (e.g., the attacker holds administrative privileges inside a guest VM)—thereby threatening co-located migrating workloads.

TALOS mitigates these threats by ensuring that orchestration of the migration process is anchored in its trusted computing base (TCB) and by enforcing continuous verification of both state and execution flow throughout the migration, thereby limiting the adversary’s ability to tamper with either the migration-control or migration-data aspects.

\subsubsection{\textbf{DDoS and Fork Bomb Attacks}}

Distributed Denial-of-Service (DDoS) attacks threaten the availability of live migration protocols by overwhelming system resources. A fork bomb—a specialized DDoS vector—spawns recursive processes, rapidly exhausting CPU and memory, leading to system-wide failure. In a migration setting, this attack can be weaponized by injecting a latent fork bomb into the application’s state during transfer. Upon launch at the destination, the payload triggers a denial-of-service from within the trusted application. CVE-2025-41227 (VMware) demonstrates how even non-privileged users in guest VMs can exhaust host memory, while CVE-2025-49722 (Windows Print Spooler) highlights the susceptibility of core services to resource exhaustion, validating the threat model.

\subsubsection{\textbf{Cloning Attacks}}

Cloning attacks occur when an adversary—typically with OS-level control—launches multiple instances of a legitimate (enclave) binary. If these instances share access to the same sealed state, application logic can be subverted. For example, a rate-limiting counter can be bypassed by distributing brute-force attempts across multiple clones. Although TEE hardware often lacks native clone prevention, vulnerabilities such as CVE-2024-21626 (container escape in \texttt{runc}) and CVE-2025-1974 (Kubernetes ingress-nginx takeover) demonstrate feasible attack vectors that grant sufficient privileges to instantiate cloned (enclave) binaries. TALOS counters this via clone-aware attestation and migration semantics, incorporating techniques inspired by CloneBuster~\cite{10.1145/3627106.3627187}.

\subsubsection{\textbf{Replay and Rollback Attacks}}

Replay and rollback attacks undermine state freshness—an essential property for secure stateful migration. Replay attacks involve reusing previously valid data to subvert enclave execution. At the network layer, this might involve retransmitting stale memory pages during a pre-copy migration. More subtly, attacks like MicroScope~\cite{8980361} exploit OS-level page fault handling to repeatedly re-execute enclave code, bypassing protocol-level defenses. Rollback attacks are a specific subclass where a valid but outdated sealed state is presented to the enclave, potentially reactivating previously patched vulnerabilities. Although hardware-based monotonic counters exist (e.g., in SGX), securely transferring their state during migration is non-trivial. TALOS addresses these risks through the use of cryptographic nonces, version tracking, and consistent migration of sealed data and counter states.

\begin{table}[t]
\centering
\resizebox{0.9\linewidth}{!}{%
\begin{tabular}{|l|p{2.3cm}|p{2.8cm}|}
\hline
\textbf{Threat Class} & \textbf{Attack Vector} & \textbf{Representative CVEs / Attacks} \\
\hline
\textbf{Software Tampering} & Compromise of migration orchestration or in-transit state & 
CVE-2024-5910 (Palo Alto), CVE-2025-22224–26 (VMware) \\
\hline
\textbf{DDoS / Fork Bomb} & Resource exhaustion via recursive process spawning or memory overload & 
CVE-2025-41227 (VMX memory), CVE-2025-49722 (Print Spooler)  \\
\hline
\textbf{Cloning Attacks} & Concurrent unauthorized enclave instantiation with shared state & 
CVE-2024-21626 (runc), CVE-2025-1974 (Kubernetes ingress-nginx) \\
\hline
\textbf{Replay / Rollback} & Reuse of stale or valid but outdated sealed state & 
MicroScope~\cite{8980361}, CVE-2025-22224 (TOCTOU) \\
\hline
\end{tabular}%
}
\caption{Summary of Key Threats and TALOS Countermeasures}
\label{tab:threat-summary}
\end{table}

\section{Related Work}

\def\nop{}

\newcommand{\cmark}{\ding{51}}%
\newcommand{\xmark}{\ding{55}}%

\begin{table}[!t]
    \centering
    \begin{tabular}{l|c|c|c|c|c|c} 
   Threat   &   \rotatebox{90}{ TALOS}  & \rotatebox{90}{\shortstack[l]{eMotion \\\cite{PARK2019173}}} & \rotatebox{90}{\shortstack[l]{MigSGX\\ \cite{inproceedings2}}} & \rotatebox{90}{\shortstack[l]{Clone\\ Buster \\ \cite{10.1145/3627106.3627187}}}& \cite{inproceedings} & \cite{inproceedings1} \\ \hline
    Fork Bombs & \cmark & \xmark  & \xmark & \cmark & \xmark & \cmark  \\
    Clone  & \cmark & \xmark  & \xmark & \cmark & \xmark  & \cmark \\
    Replay  & \cmark & \xmark  & \xmark & \xmark & \xmark  & \xmark \\
    Rollback  & \cmark & \xmark  & \xmark & \xmark & \cmark   & \xmark \\
    Integrity  & \cmark & \cmark  & \cmark & \xmark & \cmark & \cmark \\
    Confidentiality  & \cmark & \cmark  & \cmark & \xmark & \cmark  & \cmark \\ 
    \end{tabular}
           \caption{Comparison with competing approaches}
    \label{tab:comparison}
    \vspace{-0.2cm}
\end{table}

Live migration of workloads secured by hardware-assisted Trusted Execution Environments (TEEs) is increasingly relevant, particularly with the adoption of Intel SGX. Although our implementation targets SGX-protected applications running within the Gramine LibOS, TALOS is designed to be TEE-agnostic and adaptable across trusted execution frameworks. Unlike prior work that focuses on migrating full virtual machines or containers, TALOS addresses the secure and verifiable migration of enclaves at the application level—specifically capturing and preserving both persistent and volatile execution state. Table~\ref{tab:comparison} provides a comparative summary of TALOS against existing solutions, which we now discuss in detail: \textbf{(\roman{myctr})} \textbf{eMotion}~\cite{PARK2019173} introduces architectural modifications to SGX that allow virtual machine managers to securely transfer enclave pages across hosts using a trusted channel. While this ensures confidentiality and integrity of enclave memory, the required hardware changes limit deployability on commodity systems. Moreover, it lacks protections against cloning, rollback, and fork-based attacks. \stepcounter{myctr}\textbf{(\roman{myctr})} Gu et al.~\cite{inproceedings} propose a software-centric approach where enclaves implement custom logic to support state migration. This includes launch-time and runtime control mechanisms, internal threads for memory export, and cryptographic checksums for integrity. Some variants also use two-phase checkpointing for state quiescence. While quoting services and enclave self-termination are used to mitigate rollback, the architecture remains vulnerable to cloning, forking, and replay. \stepcounter{myctr}\textbf{(\roman{myctr})} Liang et al.~\cite{inproceedings1} present a framework that embeds dedicated migration libraries inside the enclave itself. These provide checkpoint and recovery via AES-GCM and reinitialize memory securely on the destination. Although this offers strong guarantees—particularly when combined with trusted counters—it introduces significant intrusiveness, requiring modifications to default enclave execution models. \stepcounter{myctr}\textbf{(\roman{myctr})} \textbf{MigSGX}~\cite{inproceedings2} integrates enclave migration within container lifecycles. It enables enclaves to autonomously persist and reload state using CPU-agnostic key exchange protocols. The design emphasizes performance, with pipelined memory dumping and restoring, but prioritizes state integrity over behavioral attestation or control-flow verification. \stepcounter{myctr}\textbf{(\roman{myctr})} \textbf{CloneBuster}~\cite{10.1145/3627106.3627187} targets clone and fork detection in adversarial settings using covert communication mechanisms—e.g., cache-based side channels—to detect concurrent enclave replicas. When coupled with hardware monotonic counters, this provides strong anti-fork assurances. Although not originally designed for migration, its techniques are relevant in scenarios where clones or fork bombs might be triggered during or post-migration.

\textbf{In summary}, TALOS is the first to unify live enclave migration with verifiable state transfer and execution continuity. It comprehensively addresses tampering, rollback, cloning, and fork-based threats by introducing a composite attestation model in which both source and destination nodes mutually verify enclave authenticity and integrity across the migration lifecycle.

\section{Requirements \& Goals}
\label{sec:requirements}

\textbf{TALOS} ensures the secure and verifiable migration of enclave-based applications across devices. This is formalized through the following system requirements:

\begin{figure*}[hbt!]
    \centering
    \includegraphics[width=0.65\textwidth]{\ifsubfolders figures/\fi 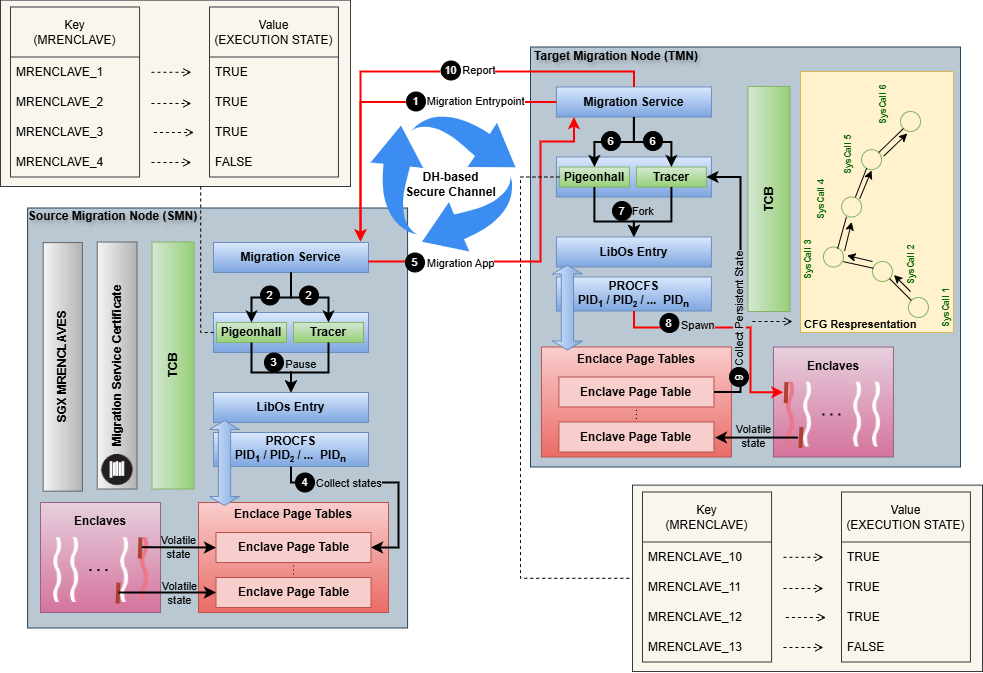}
    \caption{TALOS Conceptual Overview for Verifiable State Management During Live Migration}
    \label{fig:hlo}
\end{figure*}

\begin{enumerate}
\item[\textbf{R1}.] \textbf{Trustworthy TCB Enforcement:} Only \emph{Target} devices that provide a verifiable and trusted TCB may participate in migration. The Orchestrator specifies and validates the security posture of each Target through remote attestation mechanisms (e.g., Intel SGX, AMD SEV), ensuring that the TALOS Migration Service is correctly instantiated and unmodified. This prevents deployment to compromised environments that could undermine application security.

\item[\textbf{R2}.] \textbf{Authenticated State Transfer:} The migration protocol must detect and reject any unauthorized modifications to the application state—whether they originate from a compromised Source or Target node. This requires strong integrity guarantees through digital signatures, cryptographic hashes, and sealed storage. Upon detecting tampering, the Target must abort the process and initiate containment measures.

\item[\textbf{R3}.] \textbf{Verifiable Enclave Restoration:} Before finalizing migration, the Source must confirm that the application has been correctly launched and is operating as expected on the Target. This is achieved through remote attestation and memory validation mechanisms, which verify that enclave state has been faithfully reconstructed. This prevents premature shutdowns that could result in silent failures or inconsistent execution.

\item[\textbf{R4}.] \textbf{Replay Resistance:} To prevent adversaries from reusing previously valid migration data, each migration event must be cryptographically unique. The protocol enforces freshness via nonces, secure timestamps, and session-bound credentials, ensuring that any replayed or delayed data is immediately rejected by the Target.

\item[\textbf{R5}.] \textbf{Protection Against Resource and Cloning Attacks:} Migration endpoints must resist threats such as denial-of-service (DoS), fork bombs, rollback, and enclave cloning. This is achieved through strict resource isolation, runtime attestation, and single-instance enforcement policies. These measures prevent the execution of multiple enclave replicas and protect against state divergence, privilege escalation, and resource exhaustion attacks.
\end{enumerate}

\section{TALOS Live Migration Scheme}

The TALOS migration protocol unfolds in four distinct phases: Enrolment, Migration Affinity, Verifiable Import, and Verification.

\subsection{High-level Overview}

TALOS supports secure enclave migration through a four-phase protocol, illustrated in Figure~\ref{fig:hlo} and detailed in the following sections.

In the \emph{Enrolment Phase}, each Migration Node undergoes a one-time onboarding process. This phase provisions the node's Migration Service with migration policies and reference measurements—such as control-flow graphs for application state loading—ensuring state continuity. As this setup occurs only once per node, the actual migration remains decentralized and does not rely on ongoing Orchestrator involvement. The \emph{Migration Affinity Phase} is initiated when a Target Migration Node (TMN) requests the transfer of a Trusted Application from a Source Migration Node (SMN), typically in response to local policy enforcement. The SMN suspends execution and securely exports the application’s persistent and volatile state. During the \emph{Verifiable Import Phase}, the TMN resumes the application within its TEE and captures runtime evidence—such as system call graphs and memory mappings—used to construct an attestation report. This report serves as a real-time \textbf{Proof of Execution (PoX)} and is returned to the SMN. Finally, in the \emph{Verification Phase}, the SMN validates the attestation from the TMN. Based on the result, it either confirms successful migration and shuts down the original application, or aborts the protocol to prevent duplicate or tampered execution.

\subsection{Enrolment Phase}

The \textbf{Enrolment Phase} establishes the foundation for secure and controlled live migration by provisioning each node’s Migration Service with the necessary credentials and access control policies. Each node generates a cryptographic key pair, whose public component is sent to the Orchestrator. The Orchestrator returns a digitally signed certificate, attesting to the node’s authenticity. This lightweight, one-time setup enables mutual authentication between nodes without requiring a persistent trusted third party during live migration~\cite{doi:10.1049/cp.2019.0133}.

Alongside the certificate, each node receives from the Orchestrator a \textbf{reference system call graph} for each migration-enabled trusted application. This graph represents the expected control-flow behavior during reinitialization, particularly while restoring the application’s volatile state. During the migration, this graph is used to verify that the application resumes execution in a legitimate and unaltered way. Any deviation from this reference during runtime tracing is treated as an indicator of tampering or compromise.

Each Migration Service also initializes a \textbf{migration policy} that enforces single-instance execution. This is achieved by mapping unique enclave identities (e.g., MRENCLAVEs) to an operational status flag using a \textbf{pigeonhole-style constraint}, which prevents simultaneous instantiation of the same enclave. This policy can be \emph{static}, defined at enrolment by the Orchestrator—suitable for tightly controlled environments—or \emph{dynamic}, autonomously maintained by the node to support flexible deployments. By enforcing single-instance guarantees at the policy level, TALOS defends proactively against cloning, fork bombs, and duplicate execution without relying solely on reactive detection mechanisms~\cite{article_pig}.

\begin{algorithm}[t]
    \caption{Migration Affinity}
    \textbf{Input:}$\vPK_{TMN}$, $\vCert_{PK_{TMN}}$, $\vMRENCLAVE$, $\sigma$ \\
    \textbf{Output:}$\vPK_{\v{SMN}}$, $\vCert_{\vPK_{\v{SMN}}}$, $\vnonce$, $\vENC_{\vstate}$, $\vMAC_{\vstate}$
    \begin{algorithmic}[1]
    \raggedright
        \State $\v{Status}=\f{Search}(\v{Pigeonhall}, \vMRENCLAVE)$ \& $\fVerify(\vCert_{\v{PK_{TMN}}}, \v{PK_{TMN}}, \vPK_I)$  \& $\fVerify(\sigma, \vMRENCLAVE, \vPK_\v{TMN})$
        \State $\v{Status}\stackrel{?}{=}\c{True}$
        \State \hspace{1em} $\v{Secret} = KDF(\vPK_\v{TMN}^{\vSK_\v{SMN}})$
        \State \hspace{1em} App: $\v{Sealed}_\vstate=\f{Seal}(\vstate, \vMRSIGNER_\v{KEY})$
        \State \hspace{1em} $\v{state}=\fUnseal(\v{Sealed_{state}}, \vMRSIGNER_\v{KEY})$
        \State \hspace{1em} $\v{nonce} = \fRandom(32)$
        \State \hspace{1em} $\vMAC_\v{state} = \fHMAC(\v{state}, \v{Secret})$
        \State \hspace{1em} $\vENC_\v{state} = \fEnc(\v{state}, \v{Secret})$
        \State \hspace{1em} $\v{reference} = \fH(\v{symbols}||\v{segments}||\v{syscall_{graph}})$
    \end{algorithmic}
\end{algorithm}

\begin{algorithm}[t]
    \caption{Verifiable Import}
    \textbf{Input:}$PK_{SMN}$ $Cert_{PK_{SMN}}$ $nonce$ $ENC_{state}$ $MAC_{state}$ \\
    \textbf{Output:}$puzzle'$
    \begin{algorithmic}[1]
        \State $Status=Verify(PK_{SMN}, Cert_{PK_{SMN}}, PK_I)$
        \State $Status\stackrel{?}{=}True$
        \State \hspace{1em} $Secret = KDF(PK_{SMN}^{SK_{TMN}})$
        \State \hspace{1em} $state=DEC(ENC_{state}, Secret)$
        \State \hspace{1em} $MAC_{state}'=HMAC(state, Secret)$
        \State \hspace{1em} $MAC_{state}'\stackrel{?}{=}MAC_{state}$
        \State \hspace{2em} $Sealed_{state}=Seal(state, MRSIGNER_{KEY})$
        \State \hspace{2em} App: Launch 
        \Statex \hspace{2em} $Unsealed_{state}=Unseal(Sealed_{state}, MRSIGNER_{KEY})$
        \State \hspace{1em} $Syscall_{graph}=CFI(App)$
        \State \hspace{1em} $symbols, segments=elf\_mapping(App)$
        \State \hspace{1em} $S = H(symbols||segments||syscall_{graph})$ 
        \Statex \hspace{1em}$puzzle' = HMAC(S||nonce, Secret)$
        \State \hspace{1em} $Update(pigeonhall, MRENCLAVE)$
    \end{algorithmic}
\end{algorithm}

\subsection{Migration Affinity Phase}
\label{subsec:affinity}

The \textbf{Migration Affinity Phase} initiates when a device, referred to as the \textit{Target Migration Node (TMN)}, requests the live migration of an enclave-protected application from a \textit{Source Migration Node (SMN)}. From this point onward, we describe the implementation details specific to Intel SGX-based systems. The \textbf{TMN’s Migration Service} begins by issuing a migration challenge to the SMN. This challenge includes a digital signature over the application’s $\mathit{MRENCLAVE}$, the TMN’s public key certificate ($Cert_{PK_{TMN}}$), and its public key ($PK_{TMN}$). Upon receiving the challenge, the SMN verifies its authenticity and checks whether the requested $\mathit{MRENCLAVE}$ is currently active by querying a secure enclave registry (e.g., \textit{Pigeonhole}) \textbf{(Step~1)}. Once the challenge is validated, the two nodes establish a secure communication channel. This is achieved through an \textbf{Elliptic Curve Diffie-Hellman (ECDH)} key exchange~\cite{10.5555/3381569.3381608}, followed by a key derivation function (KDF) to produce a shared symmetric key. This key ensures confidentiality and forward secrecy for the state transfer \textbf{(Step~3)}. The SMN's Migration Service then instructs the enclave to externalize its \textbf{volatile state}, which includes heap data, stack contents, open file descriptors, and any enclave-specific secrets required for continued execution. This state is sealed using the SGX platform’s enclave-specific sealing mechanism, such as the $\texttt{MRSIGNER\_KEY}$, ensuring that it remains bound to the identity and signer of the enclave \textbf{(Step~4)}. Once sealed, the file is returned to the Migration Service and subsequently unsealed to verify its integrity and prepare it for transmission \textbf{(Step~5)}. The volatile state is then encrypted and authenticated using the derived symmetric key \textbf{(Steps~7–8)}. In parallel, the Migration Service generates a reference hash over the enclave’s \textbf{persistent state}. This includes memory mappings (e.g., \texttt{.text}, \texttt{.data}, relocation entries), program symbols, and the expected \textbf{system call graph}—previously provisioned during enrolment. This hash ensures that the control flow and memory layout of the enclave have not been tampered with \textbf{(Step~9)}. Finally, the TMN receives the following package from the SMN: the encrypted volatile state, an authentication digest, the SMN’s public key ($PK_{SMN}$), its certificate ($Cert_{PK_{SMN}}$), and a fresh nonce value. These elements are used to validate the authenticity, freshness, and integrity of the incoming application state before proceeding with enclave reinitialization.

\begin{algorithm}[t]
    \caption{Verification}
    \textbf{Input:}$\v{puzzle'}$ \\
    \textbf{Output:}$\c{Success/Fail}$
    \begin{algorithmic}[1]
        \State  $\vpuzzle=\fHMAC(\v{reference}\|| \vnonce, \v{Secret})$
        \State  $\vpuzzle'\stackrel{?}{=}\vpuzzle$
        \State \hspace{1em} $\f{Update}(\v{pigeonhall, MRENCLAVE})$
    \end{algorithmic}
\end{algorithm}

\subsection{Verifiable Import Phase}
\label{subsec:import}

In the \textbf{Verifiable Import Phase}, the Migration Service of the Target Migration Node (TMN) receives the encrypted volatile state of the trusted application and begins securely resuming its execution within the local trusted environment. To decrypt the received state, the TMN derives the shared secret by performing the Elliptic Curve Diffie-Hellman (ECDH) operation using its private key and the SMN's public key. The resulting secret is then processed through the agreed key derivation function (KDF) to produce a symmetric encryption key \textbf{(Step~3)}. This key is used to decrypt the volatile state and recompute the HMAC-SHA256 digest. A match with the received digest confirms both the authenticity and integrity of the state \textbf{(Steps~5–6)}. After decryption \textbf{(Step~4)}, the volatile state is locally sealed using the TMN’s $\mathit{MRSIGNER\_KEY}$, binding it to the enclave’s identity \textbf{(Step~7)}. The trusted application is then launched as a \textit{child enclave} of the TMN’s Migration Service \textbf{(Steps~8–9)}. This parent-child relationship allows the Migration Service to introspect the memory and execution context of the newly launched enclave without violating isolation boundaries. During runtime, the Migration Service performs introspection on the enclave’s memory layout, program symbols, and relocation segments, and simultaneously traces the control flow through its system calls. These observations verify that (i) the enclave was loaded into memory as expected, (ii) its persistent state was not tampered with, and (iii) the volatile state was successfully restored. The traced sequence of system calls acts as an implicit proof that the enclave resumed correctly from the saved state \textbf{(Steps~9–10)}, thereby ensuring execution continuity across nodes and protecting against rollback, cloning, and replay attacks described in the threat model. The collected observations and the original nonce are combined to produce a fresh attestation digest via a secure hash function \textbf{(Step~11)}. This digest is returned to the SMN as a real-time proof-of-execution (\textbf{PoX}). Finally, the TMN updates its local pigeonhole mapping with the measurement of the newly instantiated enclave ($\mathit{MRENCLAVE}$), marking it as a validated and active instance \textbf{(Step~12)}. This final step ensures global uniqueness of execution and formally concludes the Verifiable Import Phase.

\begin{figure}[!t]
    \centering
    \vspace{-0.4cm}
  \includegraphics[width=0.5\textwidth]{\ifsubfolders figures/\fi 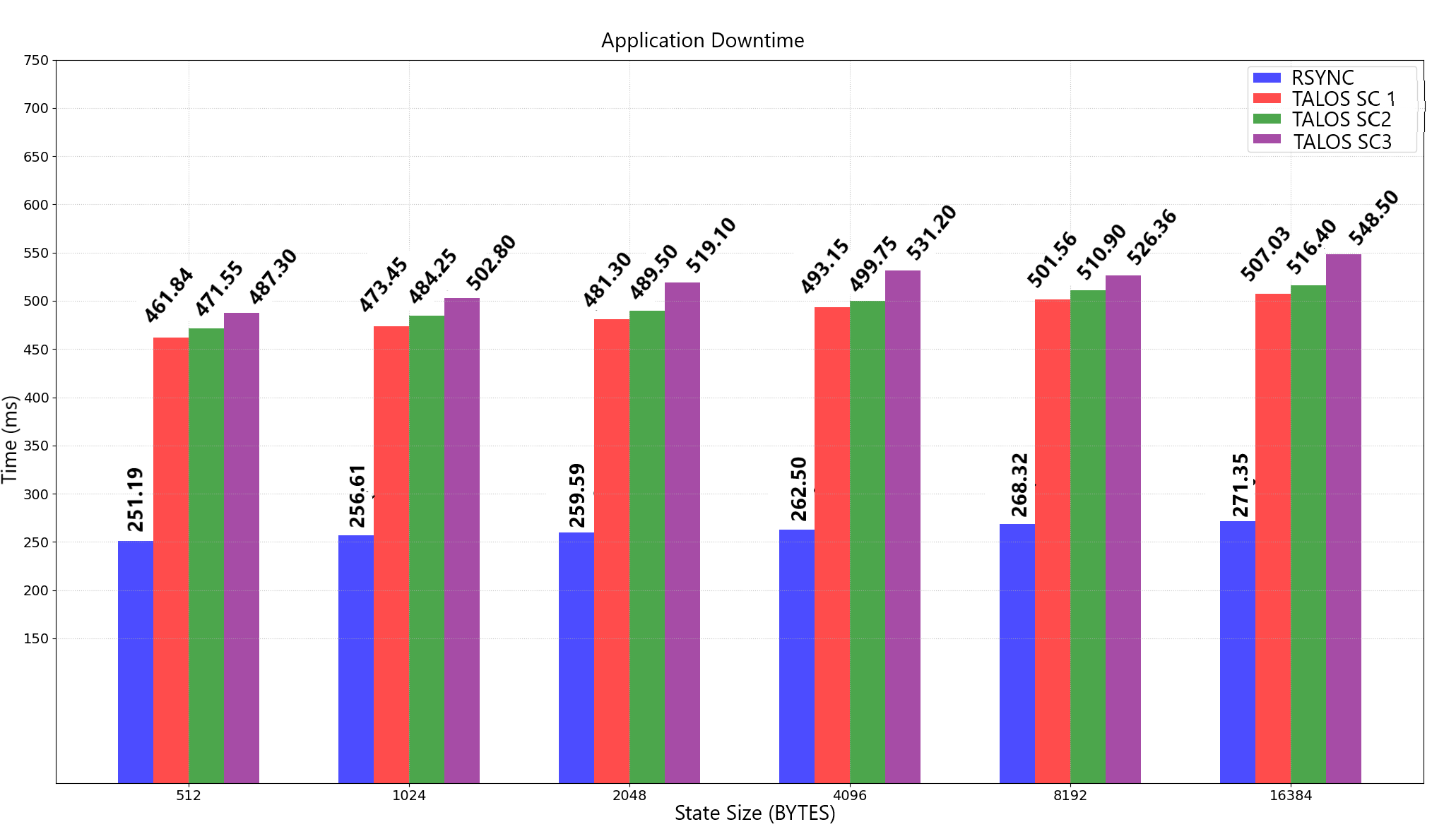}
  \caption{Average duration of (end-to-end) TALOS Secure Migration Service vs. Default RSYNC Protocol}
  \label{fig:downtime_bar}
  \vspace{-0.2cm}
\end{figure}

\subsection{Verification Phase}

The \textbf{Verification Phase} finalizes the migration protocol by validating the successful launch and execution continuity of the trusted application on the Target Migration Node (TMN). The Source Migration Node (SMN) receives the authentication digest produced during the Verifiable Import Phase. To confirm the integrity of the migration, the SMN recomputes the expected digest using its reference state hash, the nonce, and the shared secret established during the secure channel setup. A successful match confirms that the enclave has been correctly launched and that the volatile state was faithfully restored. Once verification succeeds, the SMN’s Migration Service securely terminates the original enclave instance, ensuring that only a single active copy of the application exists across the infrastructure. The SMN then updates its local pigeonhole registry to mark the enclave as successfully migrated, binding the application's identity to the TMN. This step enforces strong execution uniqueness and formally concludes the migration protocol.

\begin{table*}[hbt!]
    \centering
    \begin{tabular}{|p{0.12\linewidth}|p{0.8\linewidth}|}
    \hline
        \multicolumn{2}{|c|}{Migration Affinity} \\
         \hline 
         Verify TMN & The source migration node (SMN) verifies the certificate of the target migration node's (TMN) public key, which is signed by the orchestrator. It also confirms that the requested application is running on the SMN using the Pigeonhole principle and verifies the application's signature through its $MRENCLAVE$ value. \\   \hline
         Extract App. State & The migration node instructs the application to dump its state. This involves sealing the state with a hardware-based key to ensure its confidentiality and integrity. \\ \hline

        Mask State & The source computes a shared secret using ECDH, unseals the application state, encrypts it using the shared secret, and generates an HMAC to ensure authenticity and integrity.\\ \hline
        \multicolumn{2}{|c|}{Verifiable Import Phase} \\
         \hline
         SC-CFI & This step involves the extraction of the system call control flow to ensure runtime integrity.   \\ \hline
         UnMask State & The encrypted state is decrypted at the target using the shared secret derived from ECDH. \\ \hline
         Dump App. State & The application state is reconstructed at the target system sealed with the Intel SGX hardware based key. \\ \hline
         ELF Conf & This phase extracts and verifies the configuration of the application that has just been deployed. \\ \hline
    \end{tabular}
    \caption{TALOS Phases Breakdown}
    \label{tab:phases_breakdown}
\end{table*}

\begin{table}[hbt!]
    \centering
    \begin{tabular}{|p{0.2\linewidth}|p{0.115\linewidth}|p{0.115\linewidth}|p{0.13\linewidth}|p{0.12\linewidth}|}
    \hline
        \multicolumn{5}{|c|}{Migration Affinity: 259.28 ms} \\
         \hline 
         Subcontrol & Min(ms) & Max(ms) & Mean(ms) & std (ms) \\
         \hline
         Verify TMN & 21.36 & 43.44 & 35.12 & 6.38 \\ \hline
         Extract App. State & 32.83 & 57.55 & 41.14 & 6.47 \\ \hline
         Mask State & 132.91 & 183.64 & 164.61 & 2.61 \\ \hline         
         \multicolumn{5}{|c|}{Verifiable Import: 461.56 ms} \\
         \hline
         Verify SMN & 20.48 & 44.56 & 59.35 & 8.48 \\ \hline
         UnMask State & 135.62 & 257.12 & 168.01 & 2.51 \\ \hline
         Dump App State & 31.29 & 56.51 & 50.32 & 2.51 \\ \hline
         SC-CFI & 112.97 & 144.35 & 131.64 & 6.14 \\ \hline
         ELF Conf & 28.12 & 94.40 & 52.28 & 6.85 \\ \hline
         \multicolumn{5}{|c|}{Verification: 86.12 ms} \\
         \hline
    \end{tabular}
    \caption{TALOS Micro-benchmarks}
    \label{tab:evaluation}
    \vspace{-0.2cm}
\end{table}

\section{Implementation \& Evaluation}

This section presents the performance evaluation of \textbf{TALOS}, with an emphasis on its \textit{online commands}—namely the three core phases that constitute live migration: \textbf{Migration Affinity}, \textbf{Verifiable Import}, and \textbf{Verification}. While \textit{offline commands} such as the Enrolment Phase have also been analyzed, they are excluded from this section, as they are executed only once during the initialization of each migration node and do not impact the protocol's runtime behavior. Our primary objective is to \textbf{analytically assess the computational complexity and runtime overhead} introduced by TALOS during live migration. In particular, we examine its impact on critical system properties such as responsiveness, liveness, and migration transparency. This evaluation aims to demonstrate that TALOS enables \textbf{controlled and secure live migration with minimal disruption} to running applications. To achieve this, we measure the performance of verifiable state management operations and runtime control-flow integrity (CFI) violation detection. These metrics provide a quantitative basis for assessing the effectiveness and cost of TALOS’s monitoring mechanisms, validating its robustness as application complexity scales.

\subsection{Implementation Report}
\label{subsec:implementation-report}

The implementation of \textbf{TALOS} supports \textbf{verifiable state migration} in a flexible and lightweight manner. To achieve this, we introduced a minimal modification—just \textbf{30 lines of code}—to the standard Gramine LibOS. This small footprint preserves the minimality of the trusted computing base (TCB), ensuring ease of auditing and reinforcing TALOS’s suitability for real-world deployments without disrupting the existing LibOS architecture or ecosystem.

Importantly, TALOS is designed to operate independently of any specific \textbf{Root of Trust (RoT)} or proprietary attestation service. This independence enhances its resilience against \textbf{fragmentation in the hardware trust anchor landscape}. As long as a system supports key security primitives—namely, \textbf{enclave-based isolation, verifiable remote attestation, and sealed state protection}—TALOS can function as a \textbf{transparent enabler of verifiable state migration} with minimal system intrusion.

This design makes TALOS not only practical and easy to integrate, but also highly portable across heterogeneous platforms and trusted execution environments.

To construct the persistent state of the migrated application, the Migration Service performs a comprehensive introspection of the application's runtime within the SGX-protected environment. This reconstruction involves two key steps: extracting the application's memory layout and capturing its control-flow behavior. 

First, the Migration Service utilizes the pseudo-filesystem exposed by the Gramine LibOS, specifically \texttt{/proc/mem}, to trace and extract the memory regions allocated to the migratable application. This interface provides critical metadata, including:
\begin{itemize}
    \item Virtual memory addresses associated with the application,
    \item Segment-level permissions (read, write, execute) for each memory region,
    \item Memory-mapped files linked to the application's runtime environment.
\end{itemize}

Using this metadata, the Migration Service constructs a dynamic mapping between memory offsets, program symbols, and relocation segments. This enables real-time monitoring of the application's static memory configuration and offers fine-grained visibility into how program symbols are aligned within the enclave’s memory. This step is essential to ensure that the application’s code and data remain intact and are correctly positioned upon re-launch in the Target Migration Node (TMN).

To complement the analysis of the application's static configuration, the Migration Service also performs control-flow introspection during reinitialization. This is facilitated by a minimal extension to the Gramine LibOS: a dedicated pseudo-filesystem, \texttt{/sys\_log}, implemented as a structured worker thread responsible for capturing system call activity.

This logging mechanism enables secure enclave monitoring and supports TALOS’s role as a silent observer, allowing fine-grained introspection without interfering with application execution. Its core functionalities include:
\begin{itemize}
    \item On-demand logging of system calls, capturing both the invocation order and associated parameters;
    \item Structured and persistent storage of trace data within \texttt{/sys\_log}, retained until explicitly terminated;
    \item Reconstruction of control-flow graphs from the recorded system call sequences, enabling validation against pre-established reference behavior.
\end{itemize}

Together, the memory layout extracted from \texttt{/proc/mem} and the control-flow data collected via \texttt{/sys\_log} form the foundation of TALOS’s persistent state reconstruction. This dual-introspection process ensures that the enclave has been correctly instantiated within the Target Migration Node (TMN) and that its static and dynamic states faithfully align with the expected execution profile. In doing so, TALOS enforces state continuity and supports post-migration integrity verification.

\subsection{Too Much to Trust? Desiderata for TALOS HW-Agnostic Migration}

To demonstrate its agnostic design, \textbf{TALOS} has been prototyped on both Intel SGX with the Gramine LibOS and RISC-V Keystone, evidencing portability across heterogeneous TEE platforms. This portability is achieved by decoupling the core cryptographic protocols—largely platform-independent—from the \textbf{Proof of Execution (PoX)} mechanisms, which are inherently tied to platform-specific introspection capabilities. TALOS' cryptographic guarantees are based on common foundational features in commodity TEEs, namely hardware-bound sealing keys and remote attestation primitives. These features are available in Intel SGX, ARM TrustZone, and RISC-V Keystone, making TALOS’s cryptographic agility layer inherently portable across TEE implementations. The most significant hw-connect that might limit TALOS applicability lies in the PoX subsystem, which requires introspection of runtime memory layout and system call behavior. To accommodate this, TALOS modularizes the PoX logic, treating it as a pluggable component that can be adapted to the specific introspection APIs of each platform. This design choice ensures that while TALOS’s guarantees of confidentiality, integrity, and authenticated state continuity remain intact, only the platform-specific Tracer (PoX module) must be tailored for each TEE. To validate this design, we instantiated the PoX layer on the open-source \textbf{RISC-V Keystone} framework. Keystone enforces enclave isolation using the Physical Memory Protection (PMP) unit, managed by a machine-mode \textit{Security Monitor} (SM) \cite{keystone}. To enable TALOS introspection in this environment, we introduced two custom Supervisor Binary Interface (SBI) calls:

\begin{itemize}
    \item \textbf{Secure Memory Introspection SBI}: Allows a trusted Keystone service to extract selected memory regions from a target enclave via the SM, enabling ELF layout reconstruction—analogous to the \texttt{/proc/mem} interface in the Gramine-based prototype.

    \item \textbf{System Call Tracing Configuration SBI}: Reconfigures the enclave’s runtime to record system calls into an in-enclave pseudo-filesystem, which is sealed using a Keystone-derived key—replicating the behavior of the \texttt{/sys\_log} interface used in our SGX implementation.
\end{itemize}

These extensions demonstrate that TALOS’s PoX mechanisms are adaptable to TEEs beyond Intel SGX, thereby affirming the framework’s portability and real-world applicability across diverse trusted computing stacks.

\subsection{EXPERIMENTAL SETUP}

To assess the performance of our Gramine-based implementation, we conducted experiments on a SUPERMICRO E302 server with an Intel\textsuperscript{\textregistered} Xeon\textsuperscript{\textregistered} D-1736NT processor. Each evaluation scenario involved 1,000 test runs per application type, with average execution times recorded to capture performance trends and variability.

We structured the evaluation around two core scenarios. The first varied the volatile state size to measure how runtime dynamics—such as memory footprint and context data—affect TALOS during migration. The second varied the persistent state size, reflecting application complexity through static code, data segments, and control flow behavior that must be preserved and verified.

An integral part of our analysis is the migrated application configuration verification, which ensures the integrity and authenticity of the re-executed binary. This step underpins TALOS’s attestation mechanism, guaranteeing that only unaltered and trustworthy applications are resumed on the target system.

\begin{figure*}[ht]
    \centering
    \fbox{\begin{minipage}{0.23\textwidth}
    \scriptsize
    \textbf{Game I: Replay Attack Game}
    
    \vspace{0.4em}
    \textbf{Goal:}  The adversary attempts to reuse a previously recorded migration message to perform an unauthorized replay of an older application state.
    
    \vspace{0.4em}
    \textbf{Setup:}
    \begin{enumerate}[leftmargin=1.5em]
        \item The Source node generates a fresh $\vnonce$ for each migration attempt.
        \item The adversary records $(ENC_{\vstate}, \f{MAC}_{\vstate}, \vpuzzle')$ from a previous migration session.
        \item The adversary attempts to replay these values in a new migration session.
    \end{enumerate}
    
    \vspace{0.4em}
    \textbf{Winning Condition:}  The adversary wins if the Source node accepts the replayed migration attempt as fresh, bypassing nonce verification.
    
    \vspace{0.4em}
    \textbf{Security Guarantee:} Since $\vnonce$ is generated randomly for each migration, and $\mathrm{HMAC}(S || \vnonce, Secret)$ ensures freshness, replay attacks are mitigated unless the adversary can predict the non
    % \vspace*{1.1cm}
    \end{minipage}
     }\quad
    \fbox{\begin{minipage}{0.285\textwidth}
    \scriptsize
    \textbf{Game II: Cloning Attack Game}
    
    \vspace{0.4em}
    \textbf{Goal:}  The adversary attempts to clone a migrated application and execute it on an unauthorized enclave.
    
    \vspace{0.4em}
    \textbf{Setup:}
    \begin{enumerate}[leftmargin=1.5em]
        \item The Source node attests the Target node before migration using remote attestation. 
        \item The adversary attempts forge a valid public key certificate. 
        \item The Orchestrator distributes valid public key certificates for every valid Migration Service that can participate in the migration process
    \end{enumerate}
    
    \vspace{0.4em}
    \textbf{Winning Condition:}  The adversary wins if it can:
    \begin{enumerate} [leftmargin=1.5em]
        \item Execute the migrated application on an Migration Service that is not approved by the Orchestrator. 
        \item Forge a public key certificate such that $\mathrm{Verify}(Cert_{PK}, PK_I) = True$ for an unauthorized node. 
    \end{enumerate}
    
    \vspace{0.4em}
    \textbf{Security Guarantee:} Remote attestation relies on hardware-backed reports, making it infeasible for an adversary to forge a valid certificate without access to a trusted enclave environment.
    \end{minipage}}\quad
    \fbox{
    \begin{minipage}{0.35\textwidth}
    \scriptsize
    \textbf{Game III: Fork Bomb Prevention Game}
    
    \vspace{0.4em}
    \textbf{Goal:} The adversary attempts to exploit the migration protocol to create multiple unauthorized instances of an application, overwhelming the Target node’s resources (fork bomb attack).
    
    \vspace{0.4em}
    \textbf{Setup:}
    \begin{enumerate}[leftmargin=1.5em]
        \item The Source node executes the Migration Affinity for the unique $\mathit{MRENCLAVE}$ value of the application to be migrated.
        \item The Target node maintains a set of running enclaves and applies the pigeonhole principle to check if $\mathit{MRENCLAVE}$ already exists.
        \item The adversary attempts to trigger multiple parallel migrations of the same application.
    \end{enumerate}
    \vspace{0.4em}
    \textbf{Winning Condition:}  The adversary wins if it can:
    \begin{enumerate} [leftmargin=1.5em]
        \item Migrate multiple instances of the same enclave ($\mathit{MRENCLAVE}$) without detection.
        \item Overwhelm the Target node’s enclave execution environment, leading to a denial-of-service state.
    \end{enumerate}
    
    \vspace{0.4em}
    \textbf{Security Guarantee:} By leveraging the pigeonhole principle, the Target node ensures that only a single instance of an application with a given $\mathit{MRENCLAVE}$ is running at a time, effectively mitigating fork bomb attacks.
    \vspace*{0.1cm}
    \end{minipage}
    }

    \caption{Games formalizing the TALUS Security Analysis.}
    \label{fig:games}
\end{figure*}

\subsection{Application Complexity and Performance Impact}

We evaluated TALOS under three progressively complex migration scenarios to assess how increasing persistent state complexity affects performance. These scenarios capture varying levels of ELF structure intricacy, dynamic runtime behavior, and system interactions—all of which influence the effort required for state reconstruction and validation:

\begin{itemize}
    \item \textbf{Scenario 1:} Migrated applications are statically linked, with no shared libraries or external runtime dependencies.
    \item \textbf{Scenario 2:} Applications rely on dynamically linked libraries, introducing symbol tables, relocation entries, and a more fragmented memory layout.
    \item \textbf{Scenario 3:} The most complex case, involving system-level dependencies such as interprocess communication (e.g., pipes), shared memory segments, and open file descriptors, which extend the application state beyond enclave boundaries.
\end{itemize}

These scenarios allow us to isolate the impact of application complexity on TALOS’s performance. Tables~\ref{tab:evaluation} and~\ref{tab:phases_breakdown} break down execution time across individual migration phases. Notably, the SC-CFI component—which reconstructs a fine-grained system call control flow graph—accounts for a significant portion of the overhead. This cost is intentional, trading minimal performance for stronger attestation accuracy.

In contrast, exporting the application's volatile state imposes minimal overhead—approximately 40 ms—thanks to Intel SGX’s hardware-accelerated AES sealing. This volatile state is sealed using the enclave’s hardware-bound key (e.g., $\c{MRSIGNER\_KEY}$), then encrypted via a symmetric key derived through ECDH before transmission to the target.

The majority of the performance cost arises during the final stages of migration, specifically:
(i) the control-flow trace extraction (triggered until a predefined syscall signals that the volatile state is fully reloaded), and
(ii) the introspection and reconstruction of the persistent ELF configuration, which includes memory mappings and symbol associations.

By scaling both the volatile state size and the complexity of ELF metadata, our evaluation isolates their respective contributions to migration cost. In total, end-to-end live migration with TALOS—from pausing the application to final attestation and resume or termination—takes approximately 550 ms. This compares favorably with other secure migration frameworks such as~\cite{asokan}, which offer integrity guarantees but lack execution continuity. The additional overhead introduced by TALOS is modest, at around 50 ms, while providing stronger guarantees of state continuity and attested re-execution.

\subsection{Cross-TEE Performance: Keystone Evaluation}

We benchmarked our Keystone-based implementation of TALOS on a StarFive VisionFive 2 board to assess its performance on a lightweight, open-source TEE stack. The results, summarized in Table~\ref{tab:evaluation_riscv}, confirm that TALOS retains its low-intrusion design, acting as a \textbf{silent observer} even under resource-constrained environments.

In the most demanding test case (Scenario 3), which includes system-level dependencies and runtime interactions, the Proof-of-Execution (PoX) mechanism introduces a mere \textbf{1.57\% overhead} compared to native enclave execution. This overhead is well within acceptable bounds for safety-critical workloads and confirms the feasibility of TALOS on RISC-V hardware.

The most noticeable latency arises during memory introspection, with an average of \textbf{868.11 ms}. This is expected given (i) the modest processing capabilities of the VisionFive 2 board relative to the SGX-based server, and (ii) Keystone’s reliance on a multi-context switch mechanism (U-mode $\leftrightarrow$ S-mode $\leftrightarrow$ M-mode) for secure memory access via the Security Monitor. These four transitions inherently introduce additional latency, but they do not impact correctness or overall security.

\begin{table}[hbt!]
    \centering
    \begin{tabular}{|p{0.15\linewidth}|p{0.15\linewidth}|p{0.15\linewidth}|p{0.17\linewidth}|p{0.15\linewidth}|}
        \hline
        \textbf{Phase} & \textbf{Min (ms)} & \textbf{Max (ms)} & \textbf{Mean (ms)} & \textbf{Std (ms)} \\ \hline
        SC-CFI \% & 1.39 & 1.83 & 1.57 & 0.12 \\ \hline
        ELF Conf & 867.00 & 869.42 & 868.11 & 0.78 \\ \hline
    \end{tabular}
    \caption{TALOS RISC-V PoX Performance}
    \label{tab:evaluation_riscv}
\end{table}

This empirical evaluation on RISC-V hardware validates our design claim of TEE-agnosticism. As long as the underlying TEE provides architectural hooks for introspection—such as secure SBI calls and protected memory access—TALOS can be ported with minimal adjustments. These results affirm that TALOS can deliver secure and efficient enclave migration even in constrained, open-source environments—demonstrating its effectiveness beyond traditional, vendor-specific TEE deployments.

\subsection{TALOS vs. RSYNC: Performance \& Security Trade-offs}

We selected RSYNC for this ``head-to-head'' comparison due to its widespread adoption as a de facto standard for state replication~\cite{9155403}. To ensure fairness, we excluded any network-related delays, focusing purely on the internal execution of TALOS’s components and its verifiable state management mechanisms. As illustrated in Figure~\ref{fig:downtime_bar}, TALOS exhibits stable scaling across different migration scenarios, albeit with approximately twice the execution time of RSYNC. This overhead stems directly from TALOS’s enhanced security guarantees. Unlike RSYNC, which provides no runtime integrity or continuity protections, TALOS ensures attested, verifiable, and tamper-resistant migration. Despite this additional cost, both approaches complete migration within the same order of magnitude—milliseconds—demonstrating that TALOS remains compatible with the timing requirements of safety-critical systems. Notably, our current prototype avoids parallelizing independent stages of the protocol to capture worst-case behavior. This conservative design ensures practical deployment feasibility. However, performance can be significantly improved through parallel execution of TALOS’s independent operations, indicating that the current overhead is not a fundamental limitation of the protocol.

\section{Security Analysis}
\label{sec:security-analysis}

We analyze the security guarantees of \textbf{TALOS} with respect to the security requirements defined in Section~\ref{sec:requirements}, under the system and threat model detailed in Section~\ref{subsec:threat_model}. Our focus is on evaluating how TALOS maintains resilient migration guarantees in the presence of a strong adversary capable of compromising the host OS, hypervisor, or platform software stack, but without direct access to the hardware TEE. Specifically, we examine adversarial strategies targeting the confidentiality and integrity of the migratable enclave state, as well as state continuity violations through \emph{cloning}, \emph{forking}, or \emph{rollback} attacks. A key observation is that TALOS achieves these protections without introducing new attack surfaces within the Trusted Application. The Migration Service operates as a privileged but isolated process and does not execute within the enclave itself. All migratable operations—such as state extraction, sealing, and encrypted transmission—are cryptographically bound to the application's identity and protected via platform-provided sealing and ECDH-based masking mechanisms. In summary, TALOS preserves both the secrecy and correctness of the migrated state, while ensuring that application reinitialization occurs only on verifiably trusted destinations. This makes it secure under our defined adversary model. For a detailed breakdown of the attack surface, threat mitigation strategies, and formal security games that underpin our guarantees—including integrity, continuity, and single-instance execution—we refer the reader to Appendix~\ref{appendix:security-games}, where Figures~\ref{fig:games} and~\ref{fig:games2} illustrate our formal model and adversarial capabilities.

\begin{figure}[ht]
\centering
\fbox{\begin{minipage}{0.45\textwidth}
\scriptsize
\textbf{Game IV: Volatile State Integrity Game}

\vspace{0.4em}
\textbf{Goal:} The adversary attempts to modify the application state during migration without being detected by the target node.

\vspace{0.4em}
\textbf{Setup:}
\begin{enumerate}[leftmargin=1.5em]
    \item The challenger initializes the Source node, generating an initial application state $state$.
    \item The adversary is given access to the migration process and may attempt to modify $ENC_{\vstate}$ or $MAC_{\vstate}$.
    \item The Target node decrypts $ENC_{\vstate}$ using the shared secret and verifies $MAC_{\vstate}$.
\end{enumerate}

\vspace{0.4em}
\textbf{Winning Condition:} The adversary wins if it can modify $ENC_{\vstate}$ such that:
\begin{enumerate}[leftmargin=1.5em]
    \item $state' = DEC(ENC_{\vstate}, Secret)$ $\neq state$ (i.e., the state is modified)
    \item $MAC(state', Secret) = MAC_{\vstate}$ still holds (i.e., the modification is undetected)
\end{enumerate}

\vspace{0.4em}
\textbf{Security Guarantee:} The security of HMAC ensures that any unauthorized modification to $ENC_{\vstate}$ invalidates $MAC_{\vstate}$, preventing undetected tampering.
\label{state_integrity}
\end{minipage}}
\quad%
    \fbox{\begin{minipage}{0.45\textwidth}
    \scriptsize
     \textbf{Game V: Application Integrity Game}
     \vspace{0.4em}
     \textbf{Goal:} The adversary attempts to modify the application's configuration or disrupt the application's continuity on the Target node without being detected.
    
     \vspace{0.4em}
     \textbf{Setup:}
    \begin{enumerate}[leftmargin=1.5em]
         \item The Source node computes $reference = H(symbols || segments || syscall_{graph})$.
         \item The adversary is given access to modify the application’s ELF structure, system call graph.
         \item The Target node reconstructs its own $reference'$ and sends it to the Source node to verify it against the $reference$.
     \end{enumerate}
    
     \vspace{0.4em}
     \textbf{Winning Condition:}  The adversary wins if it can modify the application such that:
 \begin{enumerate}[leftmargin=1.5em]
         \item $reference' \neq reference$ (i.e., the application is altered)
         \item The verification step $\mathrm{HMAC}(reference || \vnonce, Secret) = puzzle'$ still holds
     \end{enumerate}
    
     \vspace{0.4em}
     \textbf{Security Guarantee:} The cryptographic hash function $H()$ ensures that even a small modification to the application state results in a completely different hash, making unauthorized changes detectable.
     \label{integrity_game}
    \end{minipage}}
    \caption{Security Games used in our Analysis (cont.)}
    \label{fig:games2}
    \end{figure}

\section{Conclusion \& Future Work}
\label{sec:conclusion}

TALOS presents the first complete framework that secures the live migration of enclaves by combining trusted execution with runtime verifiability, while preventing state duplication and ensuring execution continuity. Our core contribution is the design of a system built on modern, commodity platforms, extended with minimal modifications to support secure, fine-grained migration under a broad and realistic threat model. At the heart of TALOS is a novel Gramine LibOS extension that enables on-demand extraction of system call control-flow graphs, forming the basis for behavioral attestation during application reinitialization. This is tightly integrated with our verifiable state management layer, which produces cryptographic proofs of state integrity, confidentiality, and continuity, even in untrusted or adversarial environments. As secure computing infrastructure evolves—particularly with emerging standards for trusted path rooting and decentralized trust establishment~\cite{voit-rats-trustworthy-path-routing-11}—TALOS offers a portable, practical, and forward-compatible foundation for secure enclave migration in next-generation trusted computing stacks.

\appendix

\ifsubmissionversion
\else
% \section*{Acknowledgment}
% The authors would like to thank Dmitrii Kuwaiskii for the Gramine framework and the new migration features that he implemented to enable this research. This work was partially supported by the CONNECT Project (\url{https://horizon-connect.eu/} that is funded by the European Union under grant agreement no. 101069688. 

\section{Formal Security Model \& Games}
\label{appendix:security-games}
TALOS’s foundation of trust lies in its minimal and attested Trusted Computing Base (TCB), validated during the load-time initialization of the Migration Service (MS) using standard local and remote attestation procedures~\cite{asokan, 10.1145/3579856.3582833, 10.1145/3538969.3539004}. These mechanisms require no modification, and every component and internal thread of the MS is subject to measurement and attestation. The root of trust for reporting is anchored in the platform’s attestation key, which remains accessible only to the secure enclave responsible for signing attestation quotes. TALOS's introspection logic, in turn, serves as the core mechanism for generating trustworthy runtime measurements. Importantly, TALOS preserves the same confidentiality and integrity guarantees as a standalone, non-migratable enclave. When implemented on Intel SGX with the Gramine LibOS—as one representative platform—sensitive operations such as key generation, encryption, decryption, signature verification, and process monitoring are executed within hardware-isolated enclaves. This prevents even privileged adversaries, including a compromised OS or hypervisor, from accessing critical secrets or tampering with the MS execution state. Crucially, TALOS extends this baseline of trust by offering runtime proof-of-execution for live migration. By coupling introspection-based measurements with attestation of the MS, it provides verifiable \emph{execution continuity} of migratable workloads—even under strong adversaries, as outlined in our threat model and summarized in Table~\ref{tab:comparison}. As such, TALOS fulfills Security Requirement [R1] across a broad class of TEE implementations and deployment environments.

\subsection{\textbf{Measurement Authenticity}}

To ensure that an enclave’s state—both volatile and persistent—remains authentic and unaltered during migration, TALOS enforces a dual-measurement strategy. Independent verification of the application state (by both the source and target nodes) ensures that neither can unilaterally assert the validity of the migrated state. This mutual measurement process is essential for maintaining strong integrity guarantees and detecting tampering or replay attacks during migration. All communication between the Migration Services at the Source Migration Node (SMN) and Target Migration Node (TMN) is encrypted using symmetric keys established via a Diffie-Hellman key agreement protocol, bound to the remote attestation process. This cryptographic channel offering is twofold: First, it ensures \emph{controlled migration} by enabling both the authentication of the Source and Target migrating machines (protecting against possible impersonation attacks mounted by creating a copy of the same enclave in a non-trusted device) but also the identity verification of the migration enclave through the attestation process. After that, the identity of the Migration Enclave on the destination machine is verified and authenticated by the Migration Enclave on the source machine. This ensures that the destination Migration Enclave has a valid identity and is running on an authorized machine. Secondly, the employment of such crypto structures ensures that the volatile state transmitted during migration remains confidential and is only handled by trusted components of the TALOS architecture (e.g. MS).

Before initiating migration, the Migration Service at the SMN computes an HMAC over the application’s volatile state. This HMAC serves as a cryptographic guarantee of the integrity and authenticity of the transmitted data. Upon reception, the TMN’s Migration Service verifies this HMAC, rejecting the state if any deviation or corruption is detected.

Following successful verification, the Migration Service at the TMN performs an independent measurement over the freshly extracted persistent state of the migrated application. This involves computing a second HMAC over the persistent state together with a nonce previously issued by the SMN’s Migration Service. The use of a nonce ensures that each migration session is uniquely bound to a specific migration instance, effectively mitigating replay attacks. Matching this HMAC with the expected value confirms the integrity and authenticity of the migration process. By tightly coupling state verification to both secure communication and runtime validation, TALOS ensures that any unauthorized modification, whether during transmission or deployment, is reliably detected. This layered cryptographic defense, combined with minimal trust assumptions, underpins the robustness of TALOS in adversarial environments (ensuring Security Requirements [R2, R3], [Games IV,V] in Figure~\ref{fig:games2}).

\subsection{\textbf{Application Integrity Assurance}}

TALOS operates under a zero-trust model that deliberately avoids enforcing assumptions about the internal structure or source code of the migrated application. This choice ensures broad compatibility and flexibility across heterogeneous workloads and platforms. Instead of relying on static code validation, TALOS introduces a novel tracing-based integrity assurance mechanism—enabled by a lightweight introspection layer atop the execution environment (e.g., our implementation using Gramine LibOS). Through coordinated memory introspection and control-flow extraction—performed both before and after migration—TALOS gathers fine-grained runtime evidence that characterizes the application's operational state. This evidence includes system call traces and in-memory layout mappings that reflect the real-time behavior of the application. The introspection is performed independently by the Migration Services on both the Source and Target Migration Nodes (SMN and TMN), ensuring redundancy and mutual verification. To guarantee integrity, the collected runtime evidence is authenticated using HMACs derived from fresh nonces and a shared secret established during the migration protocol. These cryptographic bindings form the basis of TALOS's chain of trust, establishing a secure and verifiable link between the application’s pre- and post-migration state. This process enables robust detection of tampering, fork attempts, or rollback attacks—providing strong integrity guarantees even in adversarial environments. For a formal definition of this guarantee, see the security game [Game V] in Figure~\ref{fig:games2}.

\subsection{\textbf{Rollback Prevention}} On any commodity system, featuring HW-based security safeguards, rollback attacks are prevented using \emph{counters}; e.g., SGX counters~\cite{intel-counter}. These are hardware-backed (monotonic) counters that are provided to each enclave and for which the underlying security element (SGX) guarantees that they cannot be decremented. Existing (SGX-based) solutions~\cite{10.1145/3341301.3359627, 10.1145/3064176.3064213} leverage such data structures to provide a trusted counter service to persist the counter value representing the version of the host application enclave. This is done by incrementing the hardware counter and sealing the new counter value along with the enclave's state as a version number. Although this can prevent undetected replay attacks\footnote{Where an adversary saves the (encrypted) state of a system and starts  new instance using the exact same state.}, it cannot solely protect against rollback attacks if the migration mechanism cannot migrate (hardware) counters. This is what TALOS innovates through its verifiable state management and introspection: Such counter values are extracted as part of the SMN volatile state and if not correctly loaded to the destination enclave (of the TMN), the process terminates. TALOS implementation captures these values as \emph{counter offsets} which are then added to the counter value of the destination enclave (initialized to zero upon launch from the TMN Migration Service) for computing the effective counter value. The integrity and freshness of the counter value is ensured due to the TALOS masking process (Section~\ref{subsec:affinity}) and the monotonic nature of these counter data structures, thereby satisfying security requirements [R4, R5].       

\subsection{\textbf{Fork Prevention}}

To prevent resource exhaustion and unauthorized duplication of enclave instances—such as those seen in fork bomb attacks—TALOS incorporates a principled accounting mechanism inspired by the pigeonhole principle. Rather than relying on implicit trust in the host environment, TALOS enforces a strict one-to-one mapping between enclave measurements and their active execution status. This mapping is maintained by the Migration Service on the Target Migration Node (TMN), which operates a secure registry of enclave instances. Once a migration completes and the enclave is re-instantiated, the TMN’s registry is updated to reflect the new active measurement, while simultaneously marking the previous instance on the Source Migration Node (SMN) as finalized. No enclave can be re-launched unless it is uniquely associated with a valid migration token and a registry slot. By doing so, TALOS ensures that only a single valid instance of a migrated application can exist at any given time—preventing duplicated executions and limiting the attack surface against fork bombs or cloning attempts. This enforcement satisfies Security Requirement [R5] and is captured formally in [Game III] of Figure~\ref{fig:games}.

\subsection{\textbf{Cloning Prevention}}

TALOS prevents cloning attacks by embedding robust remote attestation procedures that authenticate the identity and legitimacy of migration participants. A central component of this process is a trusted Orchestrator (see Section~\ref{subsec:system_model}), which attests the integrity of each Migration Service instance and issues a signed certificate bound to its public key. During the migration handshake, the Source Migration Node (SMN) verifies the authenticity of the Target Migration Node (TMN) by validating this certificate. Migration is only permitted if the TMN presents a valid, orchestrator-signed identity, binding the enclave’s state hand-off to a verified endpoint. This ensures that even if an adversary intercepts or mimics a migration request, they cannot impersonate a legitimate TMN without forging a valid certificate—a cryptographically infeasible task under standard assumptions. By coupling state transfer with mutual attestation and orchestrator-backed authorization, TALOS guarantees that enclave state can be migrated only once and only to a verified target—thereby preventing unauthorized duplication and satisfying Security Requirement [R5]. The formal security guarantees are captured in [Game II] of Figure~\ref{fig:games}.

\subsection{\textbf{Volatile State Protection}}

The protection of the volatile state is critical to ensure seamless and secure execution continuity following migration. TALOS enforces both confidentiality and integrity through a layered cryptographic design that is portable across TEE architectures. Before transfer, the volatile state is sealed using the platform’s native hardware-backed sealing mechanism, binding the data to the source enclave's trusted identity. This ensures that the state remains confidential and tamper-evident, even if stored or transferred through untrusted components. For transmission, TALOS derives a shared symmetric key via a key exchange protocol (e.g., ECDH), using the attested public keys of the Source and Target Migration Services. The volatile state is then encrypted and authenticated using this key, ensuring that only a verified and attested Target Migration Node (TMN) can decrypt and recover the state. This guarantees end-to-end protection against interception, tampering, or replay by adversaries during migration. Through this dual-layer scheme—hardware sealing and session-based encryption—TALOS upholds Security Requirement [R3], preserving the confidentiality, integrity, and authenticity of runtime state throughout the migration process.

\bibliographystyle{plain}
\bibliography{unified}

\end{document}
\endinput
%%
%% End of file `sample-sigconf-biblatex.tex'.